\documentclass[slac_one]{revtex4}
\usepackage{graphicx}
\usepackage{fancyhdr}
\newcommand{\lsim}
{\mathrel{\raisebox{-.3em}{$\stackrel{\displaystyle <}{\sim}$}}}

\def\asymp#1%
{\mathrel{\raisebox{-.4em}{$\widetilde{\scriptstyle #1}$}}}

\def\Nequal#1%
{\mathrel{\raisebox{-.5em}{$\stackrel{=}{\scriptstyle\rm#1}$}}}
\newcommand{\dsl}[1]{\not \hspace{-0.7mm}#1}
\def\dsl{\mathpalette\make@slash}
\def\make@slash#1#2{\setbox\z@\hbox{$#1#2$}%
  \hbox to 0pt{\hss$#1/$\hss\kern-\wd0}\box0}

\def\beq{\begin{equation}}
\def\eeq{\end{equation}}
\def\beqar{\begin{eqnarray}}
\def\eeqar{\end{eqnarray}}
\def\barr#1{\begin{array}{#1}}
\def\earr{\end{array}}
\def\bfi{\begin{figure}}
\def\efi{\end{figure}}
\def\btab{\begin{table}}
\def\etab{\end{table}}
\def\bce{\begin{center}}
\def\ece{\end{center}}

\def\text{\textstyle}

\def\de{\delta}

\def\la{\lambda}

\def\si{\sigma}

\def\reffi#1{\mbox{Figure~\ref{#1}}}

\def\citere#1{\mbox{Ref.~\cite{#1}}}
\def\citeres#1{\mbox{Refs.~\cite{#1}}}

\newcommand{\TeV}{\unskip\,\mathrm{TeV}}
\newcommand{\GeV}{\unskip\,\mathrm{GeV}}
\newcommand{\MeV}{\unskip\,\mathrm{MeV}}

\newcommand{\Oa}{\mathswitch{{\cal{O}}(\alpha)}}

\def\mathswitchr#1{\relax\ifmmode{\mathrm{#1}}\else$\mathrm{#1}$\fi}

\newcommand{\PW}{\mathswitchr W}

\newcommand{\Pe}{\mathswitchr e}

\newcommand{\Pd}{\mathswitchr d}
\newcommand{\Pdbar}{\bar{\mathswitchr d}}
\newcommand{\Pu}{\mathswitchr u}

\newcommand{\Ps}{\mathswitchr s}

\newcommand{\Pc}{\mathswitchr c}

\newcommand{\Pep}{\mathswitchr {e^+}}
\newcommand{\Pem}{\mathswitchr {e^-}}
\newcommand{\PWp}{\mathswitchr {W^+}}

\def\mathswitch#1{\relax\ifmmode#1\else$#1$\fi}

\newcommand{\MW}{\mathswitch {M_\PW}}

\newcommand{\GW}{\Gamma_{\PW}}

\def\solid{\raise.9mm\hbox{\protect\rule{1.1cm}{.2mm}}}
\def\dash{\raise.9mm\hbox{\protect\rule{2mm}{.2mm}}\hspace*{1mm}}

\def\ie{i.e.\ }

\newcommand{\eefourf}{{\mathswitchr{ee4f}}}

\begin{document}

\title{{\small{2005 International Linear Collider Workshop - Stanford,
U.S.A.}}\\ 
\vspace{12pt}
Electroweak Corrections to Four-Fermion Production in
\boldmath{$\Pep\Pem$} Annihilation} 

%

\author{A.\ Denner}
\affiliation{Paul Scherrer Institut, W\"urenlingen und Villigen,
CH-5232 Villigen PSI, Switzerland}
\author{S.\ Dittmaier, M. Roth}
\affiliation{Max-Planck-Institut f\"ur Physik,
D-80805 M\"unchen, Germany}
\author{L.H.\ Wieders}
\affiliation{Paul Scherrer Institut, W\"urenlingen und Villigen,
CH-5232 Villigen PSI, Switzerland \hfill \mbox{}\\
Institute for Theoretical Physics, University of Z\"urich, CH-8057
Z\"urich, Switzerland}

\begin{abstract}
The recently completed calculation of the full
electroweak ${\cal O}(\alpha)$ corrections 
to the charged-current four-fermion production processes
$\Pep\Pem\to\nu_\tau\tau^+\mu^-\bar\nu_\mu$,
$\Pu\bar\Pd\mu^-\bar\nu_\mu$, and
$\Pu\bar\Pd\Ps\bar\Pc$ is briefly reviewed.
The calculation is performed using complex gauge-boson
masses, supplemented by complex couplings to restore gauge invariance.
The evaluation of the occurring one-loop tensor integrals,
which include 5- and 6-point functions, requires new techniques.
The effects of the complete $\Oa$ corrections to the total cross section and
to some differential cross sections of physical interest are discussed
and compared to predictions based on the double-pole approximation,
revealing that the latter approximation is not sufficient to fully
exploit the potential of a future linear collider in an analysis of
W-boson pairs at high energies.
\looseness -1
\end{abstract}

\maketitle

\thispagestyle{fancy}


\section{INTRODUCTION}

At LEP2, W-pair-mediated four-fermion $(4f)$ production was
experimentally explored with quite high precision (see \citere{lep2}
and references therein).  
The LEP2 measurements had set the scale in accuracy in the theoretical
predictions for W-pair-mediated $4f$ production, as reviewed in
\citeres{Beenakker:1996kt,Grunewald:2000ju}.  
For LEP2 accuracy, it was sufficient
to include corrections in the so-called double-pole
approximation (DPA), where only the leading term in an expansion about
the poles in the two W-boson propagators is taken into account.
Different versions of such a DPA have been used in the literature
\cite{Beenakker:1998gr,Jadach:1998tz,Jadach:2000kw,Denner:2000kn,Kurihara:2001um}.
Although several Monte Carlo programs exist that include universal
corrections, only two event generators, {\sc YFSWW}
\cite{Jadach:1998tz,Jadach:2000kw} and {\sc RacoonWW}
\cite{Denner:2000kn,Denner:1999gp,Denner:2001zp}, include
non-universal corrections.

In the DPA approach, the W-pair cross section can be predicted within
$\sim0.5\%$ $(0.7\%)$ in the energy range between $180\GeV$ ($170\GeV$)
and $\sim 500\GeV$, which was sufficient for the LEP2 accuracy of
$\sim 1\%$. In the threshold region
($\sqrt{s}\lsim 170\GeV$), the DPA is not reliable, and the best
available prediction results from an improved Born approximation (IBA)
based on leading universal corrections only, and thus possesses an
intrinsic uncertainty of $\sim 2\%$.  

At a future International $\Pe^+ \Pe^-$ Linear Collider (ILC)
\cite{Aguilar-Saavedra:2001rg,Abe:2001wn,Abe:2001gc}, the accuracy of
the cross-section measurement will be at the per-mille level, and the
precision of the $\PW$-mass determination is expected to be $\sim
10\MeV$ \cite{talkKM} by direct reconstruction and \mbox{$\sim 7\MeV$}
from a threshold scan of the total W-pair-production cross section
\cite{Aguilar-Saavedra:2001rg,Abe:2001wn}.  The theoretical
uncertainty (TU) for the direct mass reconstruction at LEP2 is
estimated to be of the order of $\sim 5\MeV$ \cite{Jadach:2001cz} to
$\lsim 10\MeV$ \cite{Cossutti}, based on results of
{\sc YFSWW} and {\sc RacoonWW}; 
theoretical improvements are, thus, desirable for an ILC.
For the cross-section prediction at threshold the TU is $\sim 2\%$,
because it is based on an IBA, and 
thus is definitely insufficient for the planned
precision measurement of $\MW$ in a threshold scan.  
The TU in constraining the anomalous triple-gauge-boson coupling
$\la_\gamma$ was
estimated to be $\sim 0.005$ \cite{Bruneliere:2002df} for the LEP2
analysis. Since a future ILC is more sensitive to anomalous 
gauge-boson couplings
than LEP2 by more than an order of magnitude, a further reduction of
the uncertainties 
resulting from missing radiative corrections is necessary. 

Recently we have completed the first 
${\cal O}(\alpha)$ calculation (improved by higher-order ISR)
for the $4f$ final states 
$\nu_\tau\tau^+\mu^-\bar\nu_\mu$,
$\Pu\bar\Pd\mu^-\bar\nu_\mu$, and
$\Pu\bar\Pd\Ps\bar\Pc$,
which are relevant for W-pair production.
We have presented results on total cross sections in
\citere{Denner:2005es} and on various differential distributions in
\citere{Denner:2005fg}. The latter publication also contains 
technical details of the actual calculation, which is rather complicated.%
\footnote{Recently the authors of the {\sc GRACE/1-LOOP} system reported on
  progress towards a full one-loop calculation for
  $\Pep\Pem\to\mu^-\bar\nu_\mu\Pu\bar\Pd$ in \citere{Boudjema:2004id}
  so that one can expect that the system will be able to deal with
  $\Pep\Pem\to4f$ processes at one loop in the near future.}
In the following we briefly describe the salient features of the
calculation and present a selection of numerical results that are
relevant for a future ILC, 
comprising total cross sections and some phenomenologically
interesting distributions.

\section{METHOD OF CALCULATION}

The actual calculation builds upon the {\sc RacoonWW} approach
\cite{Denner:2000kn}, where real-photonic corrections are based on
full matrix elements and virtual corrections are treated in DPA. Real
and virtual corrections are combined either using two-cutoff
phase-space slicing or employing the dipole subtraction method
\cite{Dittmaier:2000mb} for photon radiation.  
We also include leading-logarithmic ISR
beyond $\Oa$ in the structure-function approach
(\citere{Beenakker:1996kt} and references therein). 

\subsection{Technical issues}

In contrast to the DPA approach, the one-loop calculation of
an $\Pep\Pem\to4f$ process requires the evaluation of 5- and 6-point
one-loop tensor integrals. 
For the generic $f_1 \bar f_2 f_3 \bar f_4$ final state, where $f_1$
and $f_3$ are different fermions excluding electrons and electron
neutrinos and $f_2$ and $f_4$ their isospin partners, there are 40
hexagon diagrams, 112 pentagon diagrams, and 227 (220) box diagrams in
the conventional 't Hooft--Feynman gauge (background-field gauge).
A survey of Feynman diagrams can be found in \citere{Denner:2005fg}. 
We calculate the 6-point integrals by directly reducing them to six
5-point functions, as described in \citeres{Me65,Denner:1993kt}.  The
5-point integrals are reduced to five 4-point functions following the
method of \citere{Denner:2002ii}. Note that this
reduction of 5- and 6-point integrals to 4-point integrals does not
involve inverse Gram determinants composed of external momenta, which
naturally occur in the Passarino--Veltman reduction
\cite{Passarino:1979jh} of tensor to scalar integrals. The latter
procedure leads to serious numerical problems when the Gram
determinants become small, which happens usually near the boundary of
phase space but can also occur within phase space because of the
indefinite Minkowski metric.

We use, however, Passarino--Veltman reduction to calculate tensor
integrals up to 4-point functions, which involves inverse Gram
determinants built from two or three momenta.  This, in fact, leads to
numerical instabilities in phase-space regions where these Gram
determinants become small.  For these regions we have worked out two
``rescue systems'': one makes use of expansions of the tensor
coefficients about the limit of vanishing Gram determinants; in the
other, alternative method we numerically evaluate a specific tensor
coefficient, the integrand of which is logarithmic
in Feynman parametrization, and derive the remaining
coefficients as well as the scalar integral from it algebraically.
This reduction does not involve inverse Gram determinants.

In addition to the evaluation of the one-loop integrals, also the
evaluation of the three spinor chains corresponding to the
three external fermion--antifermion pairs is non-trivial, because
the chains are contracted with each other and/or with 
four-momenta in many different ways. There are ${\cal O}(10^3)$
different chains to calculate, so that an algebraic reduction to a
standard form which involves only very few standard chains is
desirable.  We have worked out algorithms that reduce all
occurring spinor chains to ${\cal O}(10)$ standard structures without
introducing coefficients that lead to numerical problems.
These algorithms are described in \citere{Denner:2005fg} in detail.

\subsection{Conceptual issues}

The description of resonances in (standard) perturbation theory 
requires a Dyson summation of self-energy insertions in the resonant
propagator in order to introduce the imaginary part provided by the
finite decay width into the propagator denominator.
This procedure in general violates gauge
invariance, \ie destroys Slavnov--Taylor or Ward identities
and disturbs the cancellation of gauge-parameter dependences,
because different perturbative orders are mixed
(see, for instance, \citere{Grunewald:2000ju} and references therein).
The DPA provides a gauge-invariant answer in terms of an expansion about the
resonance,
but in the full calculation we are
after a unified description that is valid both for resonant and
non-resonant regions in phase space, without any matching between
different treatments for different regions.
\looseness -1

For our calculation we have generalized \cite{Denner:2005fg}
the so-called ``complex-mass scheme'', which was introduced in
\citere{Denner:1999gp} for lowest-order calculations, to the one-loop
level.  In this approach the W- and Z-boson masses are consistently
considered as complex quantities, defined as the locations of the
poles in the complex $p^2$ plane of the corresponding propagators with
momentum $p$.  Gauge invariance is preserved if the complex masses are
introduced everywhere in the Feynman rules, in particular, in the
definition of the weak mixing angle,
which is now derived from the ratio of the complex masses.
The (algebraic) relations, such as Ward identities, that follow from gauge
invariance remain valid, because the gauge-boson masses are modified
only by an analytic continuation. Since this continuation already
modifies the lowest-order predictions by changing the gauge-boson
masses, double-counting of higher-order effects has to be carefully
avoided by an appropriate renormalization procedure.

\begin{sloppypar}
The use of complex gauge-boson masses necessitates the consistent use
of these complex masses also in loop integrals. To this end, we have
derived all relevant one-loop integrals with complex internal masses.
The IR-singular integrals were taken from \citere{Beenakker:1990jr}.
Concerning the non-IR singular cases, we have analytically continued
the results of \citere{'tHooft:1979xw} for the 2-point and 3-point
functions, and the relevant results of
\citere{Denner:1991qq} for the 4-point functions. We have checked all
these results by independent direct calculation of the
Feynman-parameter integrals.
\end{sloppypar}

\subsection{Checks on the calculation}

In order to prove the reliability of our results, we have carried out a
number of checks, as described in more detail in
\citere{Denner:2005es}.  We have checked the structure of the (UV,
soft, and collinear) singularities, the matching between virtual and
real corrections, and the gauge independence (by performing the
calculation in 't~Hooft--Feynman gauge and in the
background-field gauge \cite{Denner:1994xt}). The most convincing check for
ourselves is the fact that we worked out the whole calculation in two
independent ways, resulting in two independent computer codes the
results of which are in good agreement. All algebraic manipulations,
including the generation of Feynman diagrams, have been done using
independent programs.
{The amplitudes are generated with {\sc FeynArts}, using
  the two independent versions 1 and 3, as described in
  \citeres{Kublbeck:1990xc} and \cite{Hahn:2000kx}, respectively.
  The algebraic manipulations are performed using two independent
  in-house programs implemented in {\sc Mathematica}, one of which
  builds upon {\sc FormCalc}~\cite{Hahn:1998yk}.} 
For the calculation of the loop integrals
we use the two independent in-house libraries which employ the
different calculational methods sketched above
for the numerical stabilization.

\section{NUMERICAL RESULTS}

The precisely defined input for the numerical results presented in the
following can be found in \citeres{Denner:2005es,Denner:2005fg}.

\begin{figure}
{\unitlength 1cm
\begin{picture}(16,15.5)
\put(-4.6,- 7.2){\includegraphics{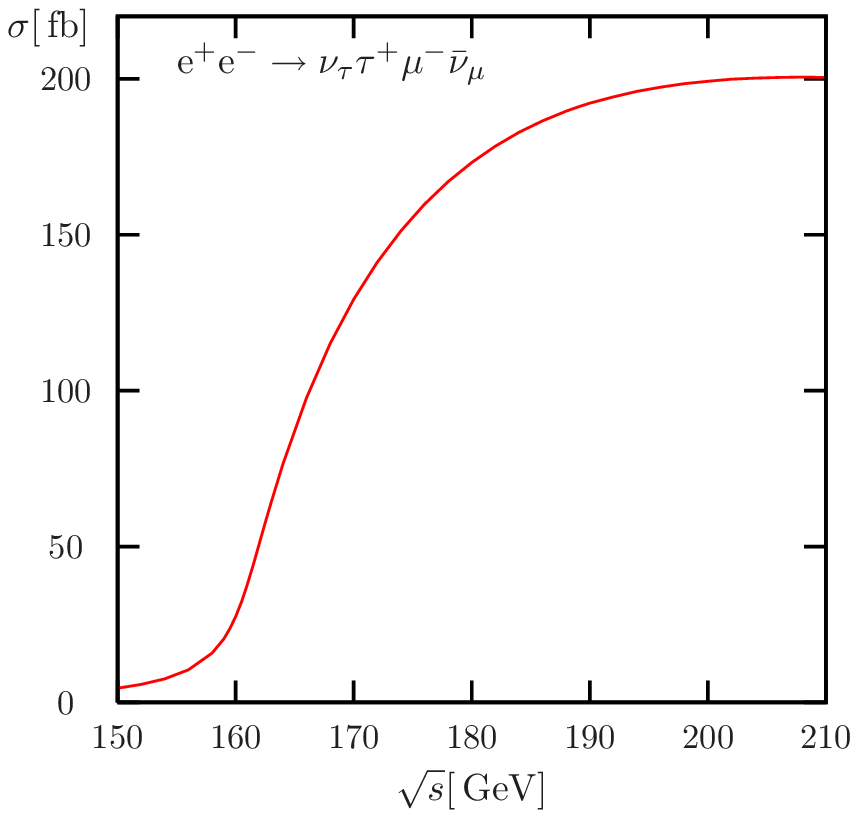}}
\put( 3.6,- 7.2){\includegraphics{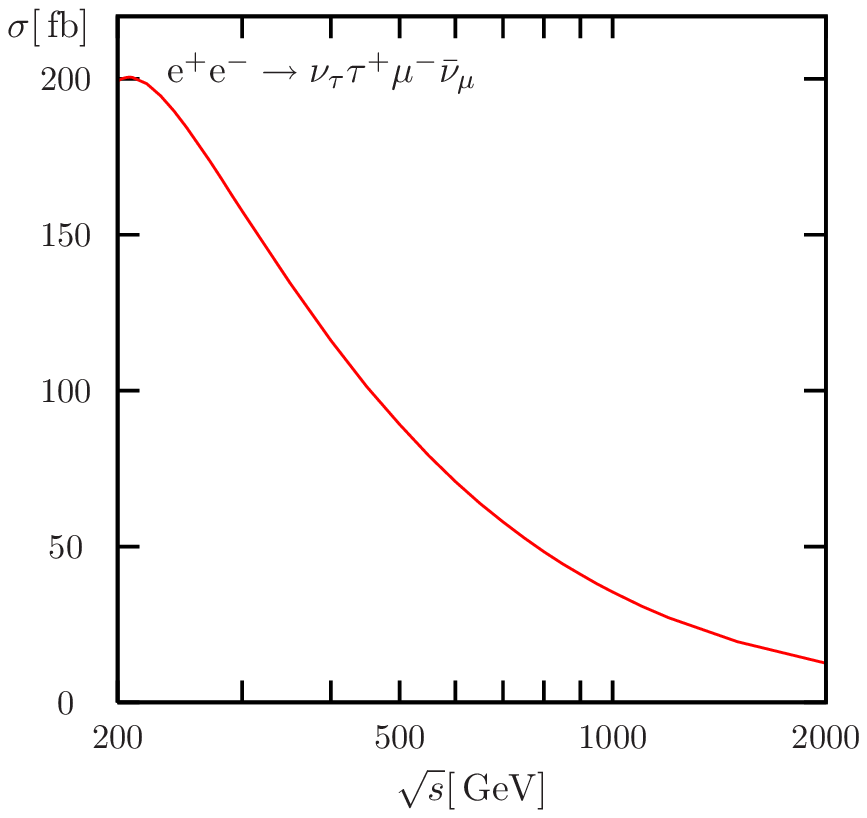}}
\put(-4.6,-15.2){\includegraphics{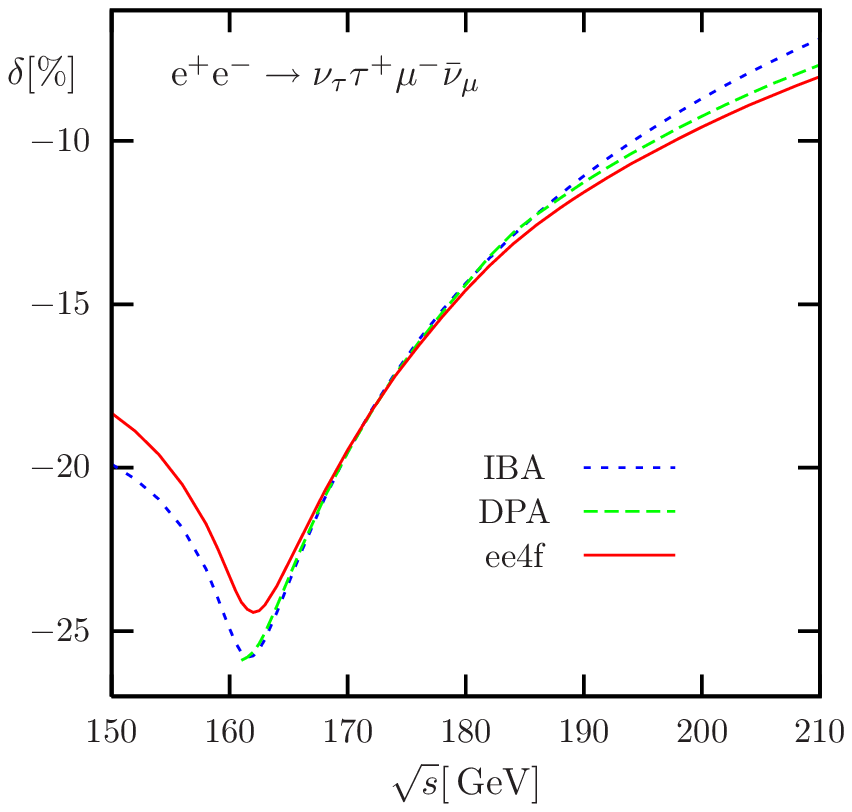}}
\put( 3.6,-15.2){\includegraphics{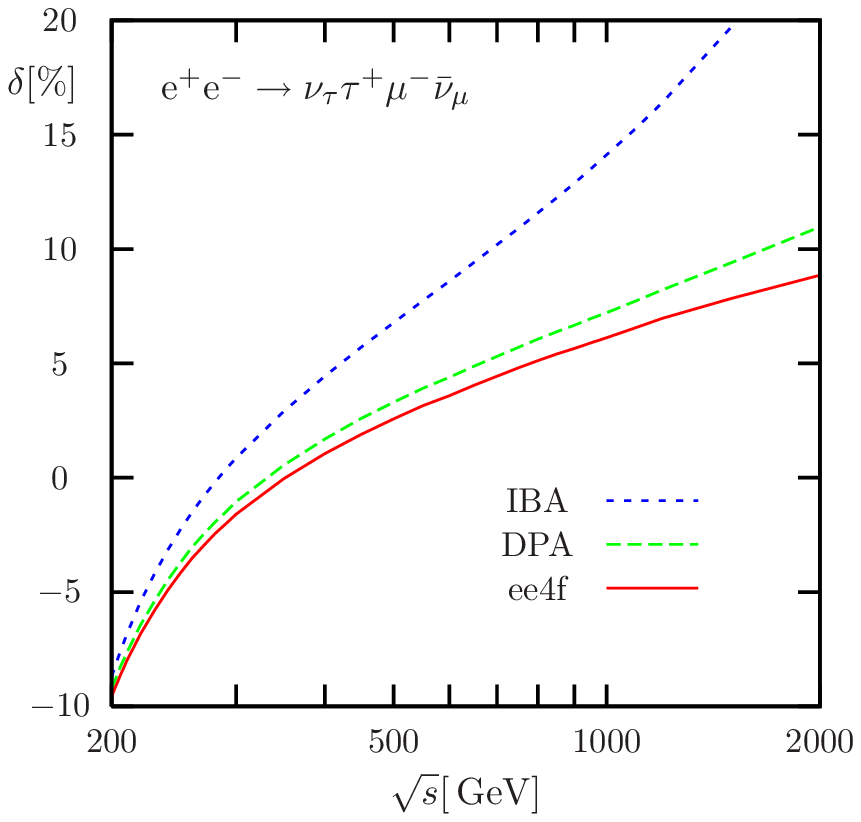}}
\end{picture} }
\caption{Absolute cross section $\si$
(upper plots) and relative corrections $\de$ (lower plots)
to the total cross section without cuts for
$\Pep\Pem\to\nu_\tau\tau^+\mu^-\bar\nu_\mu$ obtained from the
IBA, DPA, and the full ${\cal O}(\alpha)$ calculation (\eefourf). All
predictions are improved by higher-order ISR. 
(Taken from \citere{Denner:2005es}.)}
\label{fig:leptrelrc}
\end{figure}

Figure~\ref{fig:leptrelrc} depicts the total cross section
for the energy ranges of LEP2 and of the high-energy phase of a future ILC,
focusing on the leptonic final state $\nu_\tau\tau^+\mu^-\bar\nu_\mu$.
The respective figures for the relative corrections $\de$ to the
semileptonic and hadronic final states look almost identical, up to an
offset resulting from the QCD corrections.
Specifically, the upper plots show the absolute prediction for the cross 
section including the full $\Oa$ corrections and improvements from higher-order
ISR.
The lower plots compare the relative corrections as obtained from the
full $\Oa$ calculation, from an IBA, and from the DPA.
The IBA \cite{Denner:2001zp} implemented in {\sc RacoonWW} is
based on universal corrections only and includes solely the
contributions of the CC03 diagrams.
The DPA of {\sc RacoonWW}
comprises also non-universal corrections \cite{Denner:2000kn}
and goes beyond a pure
pole approximation in two respects. The real-photonic corrections
are based on the full $\Pep\Pem\to4f+\gamma$ matrix elements, and the
Coulomb singularity is included for off-shell W~bosons. 
Further details can be found in \citere{Denner:2000kn}.

A comparison between the DPA and the predictions based on the full
${\cal O}(\alpha)$ corrections reveals differences in the relative
corrections $\de$ of $\lsim 0.5\%\; (0.7\%)$ for CM energies ranging from
$\sqrt{s}\sim170\GeV$ to $300\GeV\; (500\GeV)$.  This is in agreement
with the expected reliability of the DPA, as
discussed in \citeres{Grunewald:2000ju,Jadach:2000kw,Denner:2000kn}.
At higher energies, the deviations increase and reach $1{-}2\%$ at
$\sqrt{s}=1{-}2\TeV$.
In the threshold region ($\sqrt{s}\lsim170\GeV$), as expected, the DPA
also becomes worse w.r.t.\ the full one-loop calculation, because the
naive error estimate of $({\alpha}/{\pi})\times({\GW}/{\MW})$ times
some numerical safety factor of ${\cal O}(1{-}10)$ for the corrections
missing in the DPA has to be replaced by 
$({\alpha}/{\pi})\times{\GW}/{(\sqrt{s}-2\MW})$ in the threshold region
and thus becomes large.
In view of that, the DPA is even surprisingly good near threshold.
For CM energies below $170\GeV$ the LEP2 cross section analysis was
based on approximations like the shown IBA, which follows the full
one-loop corrections even below the threshold at $\sqrt{s}=2\MW$
within an accuracy of about $2\%$, as expected in
\citere{Denner:2001zp}.
More results on total cross sections, including numbers on
leptonic, semileptonic, and hadronic final states, can be found in
\citere{Denner:2005es}.
\looseness -1

\begin{figure}
{\unitlength 1cm
\begin{picture}(16,16)
\put(-4.9,- 9.7){\includegraphics{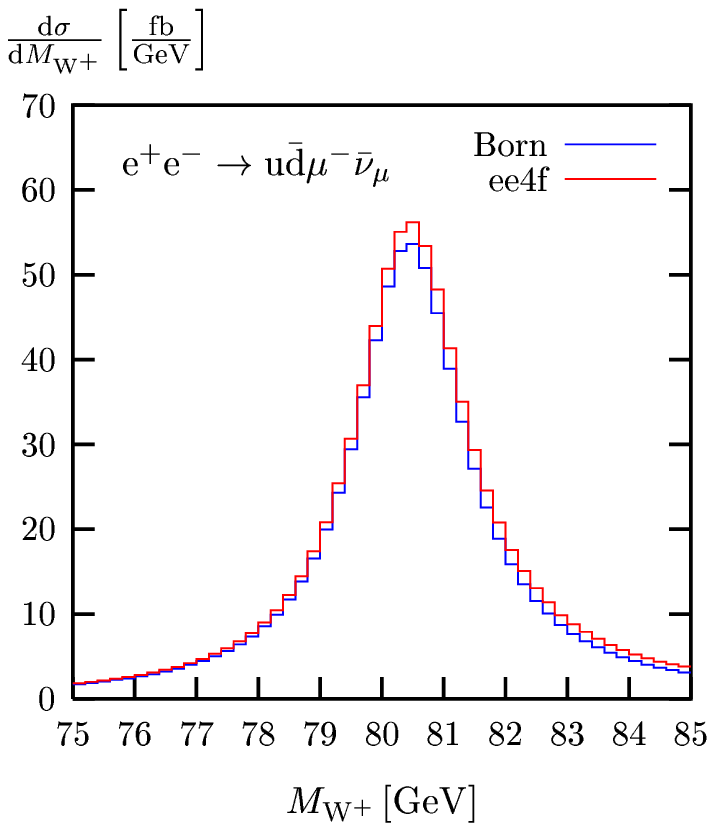}}
\put( 3.3,- 9.7){\includegraphics{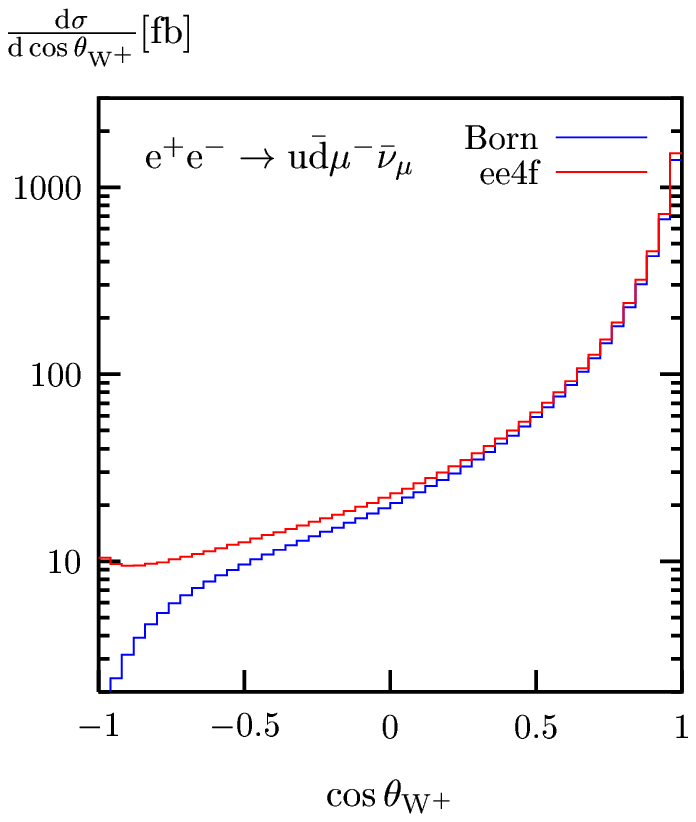}}
\put(-4.9,-17.7){\includegraphics{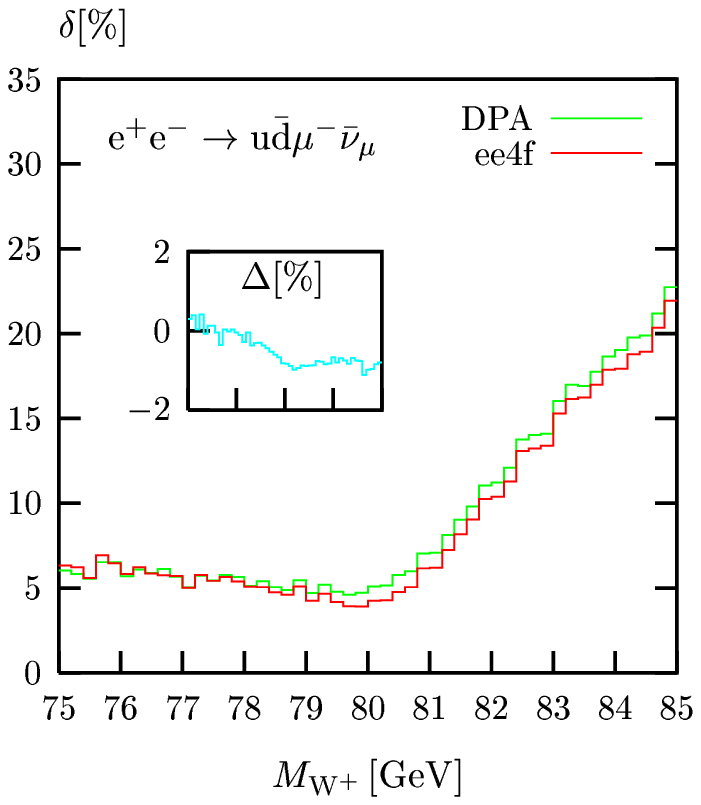}}
\put( 3.3,-17.7){\includegraphics{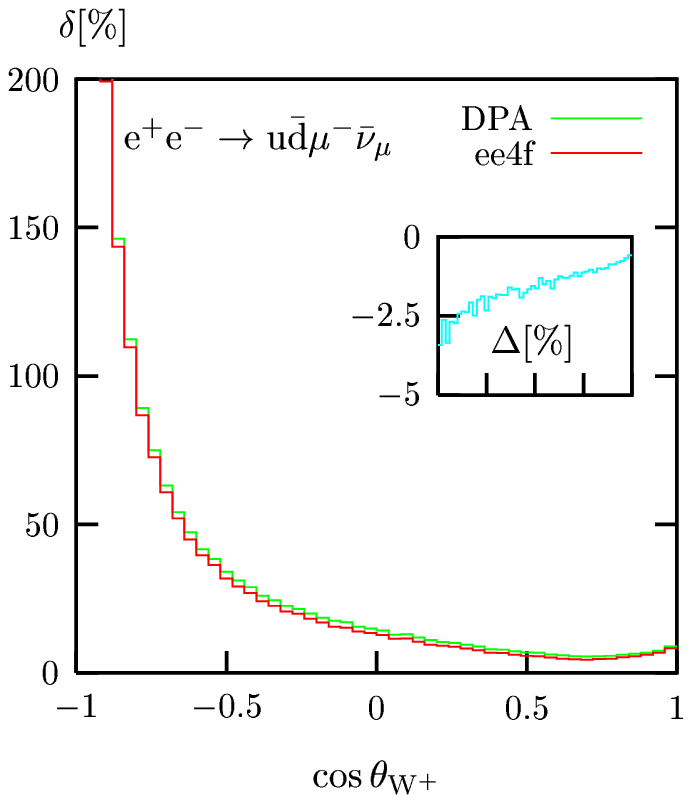}}
\put(3,15.5){$\sqrt{s}=500\GeV$}
\put(11.2,15.5){$\sqrt{s}=500\GeV$}
\end{picture}}%
\caption{Distributions in the invariant mass of the $\protect\PWp$
  boson (l.h.s.) and in the cosine of the  $\protect\PWp$ production angle
  with respect to the $\protect\Pep$ beam (r.h.s.) and the
  corresponding corrections (lower row) at $\sqrt{s}=500\GeV$ 
  for $\Pep\Pem\to \Pu \Pdbar \mu^-
  \bar\nu_{\mu}$. The inset plot shows the relative difference $\Delta$
  between the full
  $\Oa$ corrections and those in DPA. (Taken from \citere{Denner:2005fg}.)}
\label{fi:dist}
\end{figure}

The distributions in the invariant mass of the $\protect\PWp$
boson and in the cosine of the $\PWp$ production angle
$\theta_{\PWp}$ are shown in \reffi{fi:dist} for the process
$\Pep\Pem\to \Pu \Pdbar \mu^- \bar\nu_{\mu}$ at $\sqrt{s}=500\GeV$.
Further distributions, also for $\sqrt{s}=200\GeV$, are presented in
\citere{Denner:2005fg}.
For the invariant-mass distribution the full $\Oa$ calculation and the
DPA agree within $\sim 1\%$ both for LEP2 and ILC energies.
For the W-production-angle distribution this is also the case in the
LEP2 range (see Fig.~12 of \citere{Denner:2005fg}), but
at $500\GeV$ the difference of the
corrections in DPA and the complete $\Oa$ corrections rises from
$-1\%$ to about $-2.5\%$ with increasing scattering angle. Note that
such a distortion of the shape of the angular distribution can be a
signal for anomalous triple gauge-boson couplings.

\begin{acknowledgments}
This work was supported in part by the Swiss National Science
Foundation.
\end{acknowledgments}



\begin{thebibliography}{99} 


\bibitem{lep2}
The LEP Collaborations ALEPH, DELPHI, L3, OPAL, the LEP EWWG,
and the SLD Heavy Flavor and Electroweak Groups,
hep-ex/0412015.

\bibitem{Beenakker:1996kt}
W.~Beenakker {\it et al.},
in {\sl Physics at LEP2}, eds.\ G.~Altarelli, T.~Sj\"o\-strand and
F.~Zwirner (CERN 96-01, Geneva, 1996), Vol.~1, p.~79
[hep-ph/9602351].

\bibitem{Grunewald:2000ju}
M.~W.~Gr\"unewald {\it et al.},
in {\it Reports of the Working Groups on Precision Calculations
for LEP2 Physics}, eds.\ S.~Jadach, G.~Passarino and R.~Pittau
(CERN 2000-009, Geneva, 2000), p.~1
[hep-ph/0005309].

\bibitem{Beenakker:1998gr}
W.~Beenakker, F.~A.~Berends and A.~P.~Chapovsky,
Nucl.\ Phys.\ B {\bf 548}, 3 (1999)
[hep-ph/9811481].

\bibitem{Jadach:1998tz}
S.~Jadach {\it et al.},
Phys.\ Rev.\ D {\bf 61} (2000) 113010
[hep-ph/9907436];
%
Comput.\ Phys.\ Commun.\  {\bf 140} (2001) 432
[hep-ph/0103163] and
Comput.\ Phys.\ Commun.\  {\bf 140} (2001) 475
[hep-ph/0104049].

\bibitem{Jadach:2000kw}
S.~Jadach {\it et al.},
Phys.\ Rev.\ D {\bf 65} (2002) 093010
[hep-ph/0007012].

\bibitem{Denner:2000kn}
A.~Denner, S.~Dittmaier, M.~Roth and D.~Wackeroth,
Phys.\ Lett.\ B {\bf 475} (2000) 127
[hep-ph/9912261];
%
Eur.\ Phys.\ J.\ direct C {\bf 2} (2000) 4
[hep-ph/9912447];
%
Nucl.\ Phys.\ B {\bf 587} (2000) 67
[hep-ph/0006307] and
%
Comput.\ Phys.\ Commun.\  {\bf 153} (2003) 462
[hep-ph/0209330].

\bibitem{Kurihara:2001um}
Y.~Kurihara, M.~Kuroda and D.~Schildknecht,
Phys.\ Lett.\ B {\bf 509} (2001) 87
[hep-ph/0104201].

\bibitem{Denner:1999gp}
A.~Denner, S.~Dittmaier, M.~Roth and D.~Wackeroth,
Nucl.\ Phys.\ B {\bf 560} (1999) 33
[hep-ph/9904472].

\bibitem{Denner:2001zp}
A.~Denner, S.~Dittmaier, M.~Roth and D.~Wackeroth,
in {\it Proc. of the 5th International Symposium on Radiative Corrections (RADCOR 2000)} ed. H.~E. Haber,
hep-ph/0101257.


\bibitem{Aguilar-Saavedra:2001rg}
J.~A.~Aguilar-Saavedra {\it et al.},
TESLA TDR Part III: Physics at an $\mathrm{e^+e^-}$
Linear Collider,
hep-ph/0106315.

\bibitem{Abe:2001wn}
T.~Abe {\it et al.}  [American Linear Collider Working Group Collaboration],
in {\it Proc. of the APS/DPF/DPB Summer Study on the Future of
  Particle Physics (Snowmass 2001) } ed. R.~Davidson and C.~Quigg,
SLAC-R-570 {\it Resource book for Snowmass 2001},
[hep-ex/0106055, hep-ex/0106056, hep-ex/0106057, hep-ex/0106058].

\bibitem{Abe:2001gc}
K.~Abe {\it et al.}  [ACFA Linear Collider Working Group Collaboration],
ACFA Linear Collider Working Group report,
[hep-ph/0109166].

\bibitem{talkKM}
K.~M\"onig and A.~Tonazzo,
talk given by K.~M\"onig at the {\it 2nd ECFA/DESY Study on Physics and
Detectors for a Linear Electron--Positron Collider},
Padova, Italy, 2000.

\bibitem{Jadach:2001cz}
S.~Jadach {\it et al.},
Phys.\ Lett.\ B {\bf 523} (2001) 117
[hep-ph/0109072].

\bibitem{Cossutti}
F.~Cossutti, DELPHI note 2004-050 PHYS 944.

\bibitem{Bruneliere:2002df}
R.~Bruneli\`ere {\it et al.},
Phys.\ Lett.\ B {\bf 533} (2002) 75
[hep-ph/0201304].

\bibitem{Boudjema:2004id}
F.~Boudjema {\it et al.},
Nucl.\ Phys.\ Proc.\ Suppl.\  {\bf 135} (2004) 323
[hep-ph/0407079].

\bibitem{Denner:2005es}
  A.~Denner, S.~Dittmaier, M.~Roth and L.~H.~Wieders,
  Phys.\ Lett.\ B {\bf 612} (2005) 223
  [hep-ph/0502063].

\bibitem{Denner:2005fg}
  A.~Denner, S.~Dittmaier, M.~Roth and L.~H.~Wieders,
  hep-ph/0505042.

\bibitem{Dittmaier:2000mb}
S.~Dittmaier,
Nucl.\ Phys.\ B {\bf 565} (2000) 69
[hep-ph/9904440];\\
%
M.~Roth,
PhD thesis, ETH Z\"urich No. 13363 (1999),
hep-ph/0008033.

\bibitem{Me65} D.~B.~Melrose, {Nuovo Cimento}~{\bf XL~A} (1965) 181.

\bibitem{Denner:1993kt}
A.~Denner,
Fortsch.\ Phys.\  {\bf 41} (1993) 307.

\bibitem{Denner:2002ii}
A.~Denner and S.~Dittmaier,
Nucl.\ Phys.\ B {\bf 658} (2003) 175
[hep-ph/0212259].

\bibitem{Passarino:1979jh}
G.~Passarino and M.~Veltman,
Nucl.\ Phys.\ B {\bf 160} (1979) 151.

\bibitem{Beenakker:1990jr}
W.~Beenakker and A.~Denner,
Nucl.\ Phys.\ B {\bf 338} (1990) 349.

\bibitem{'tHooft:1979xw}
G.~'t Hooft and M.~Veltman,
Nucl.\ Phys.\ B {\bf 153} (1979) 365.

\bibitem{Denner:1991qq}
A.~Denner, U.~Nierste and R.~Scharf,
Nucl.\ Phys.\ B {\bf 367} (1991) 637.

\bibitem{Denner:1994xt}
A.~Denner, S.~Dittmaier and G.~Weiglein,
Nucl.\ Phys.\ B {\bf 440} (1995) 95
[hep-ph/9410338].

\bibitem{Kublbeck:1990xc}
J.~K\"ublbeck, M.~B\"ohm and A.~Denner,
Comput.\ Phys.\ Commun.\  {\bf 60} (1990) 165;\\
H.~Eck and J.~K\"ublbeck, {\it Guide to FeynArts 1.0\/},
University of W\"urzburg, 1992.

\bibitem{Hahn:2000kx}
T.~Hahn,
Comput.\ Phys.\ Commun.\  {\bf 140} (2001) 418
[hep-ph/0012260].

\bibitem{Hahn:1998yk}
T.~Hahn and M.~Perez-Victoria,
Comput.\ Phys.\ Commun.\  {\bf 118} (1999) 153
[hep-ph/9807565];\\
%
T.~Hahn,
Nucl.\ Phys.\ Proc.\ Suppl.\  {\bf 89} (2000) 231
[hep-ph/0005029].
\end{thebibliography}
\end{document}